\newif\iffigs\figstrue
\documentclass[12pt]{article}
\usepackage{latexsym,amssymb}
  \iffigs
   \input{epsf}
\else
\fi

\newcommand{\eq}{\begin{equation}}
\newcommand{\eqa}{\begin{eqnarray}}
\newcommand{\en}{\end{equation}}
\newcommand{\ena}{\end{eqnarray}}
\newcommand{\enn}{\nonumber \end{equation}}

\def\sk{\vskip .4cm}
\def\noi{\noindent}

\def\Om{\Omega}
\def\al{\alpha}
\def\la{\lambda}
\def\be{\beta}
\def\ga{\gamma}
\def\Ga{\Gamma}

\def\Cb{\bar{C}}

\def\epsi{\varepsilon}
\def\we{\wedge}

\def\de{\delta}

\def\part{\partial}

\def\R#1#2{ R^{#1}_{~~~#2} }

\def\Rb{{\bf \mbox{\boldmath $R$}}}

\def\Linv#1#2{ \La^{-1~#1}_{~~~~~#2} }

\def\Cb{{\bf \mbox{\boldmath $C$}}}

\def\n2{{{N+1} \over 2}}

\def\square{{\,\lower0.9pt\vbox{\hrule \hbox{\vrule height 0.2 cm
\hskip 0.2 cm \vrule height 0.2 cm}\hrule}\,}}

\def\Q.E.D.{\rightline{$\Box$}}

\def\Linv{L^{-1}}

\def\udot#1{\bar{#1}}

\begin{document}
\begin{titlepage}
\vskip -1cm \rightline{DFTT-74/99}
\rightline{December 1999} \vskip 1em
\begin{center}
{\large\bf On $G/H$ geometry and its use in
$M$-theory compactifications }
\\[2em]
Leonardo Castellani \\[.4em] {\sl Dipartimento di Scienze e
Tecnologie Avanzate,
 East Piedmont University, Italy; \\
Dipartimento di Fisica Teorica and Istituto Nazionale di Fisica
Nucleare\\ Via P. Giuria 1, 10125 Torino, Italy.} \\[2em]
\end{center}
\vskip 4 cm
\begin{abstract}
The Riemannian geometry of coset spaces is reviewed, with
emphasis on its applications to supergravity and
$M$-theory compactifications. Formulae for the connection and
curvature of rescaled coset manifolds are generalized to the case
of nondiagonal Killing metrics.

 The example of the $N^{010}$ spaces is discussed in detail.
 These are a subclass of the coset manifolds
 $N^{pqr}=G/H = SU(3) \times U(1)/U(1) \times U(1)$,
the integers $p,q,r$ characterizing the embedding of $H$ in $G$.
 We study the realization of $N^{010}$ as
 $G/H=SU(3) \times SU(2)/U(1) \times SU(2)$ (with diagonal embedding
 of the $SU(2) \in H$ into $G$). For a particular $G$-symmetric
 rescaling there exist three Killing spinors, implying $N=3$
 supersymmetry in the $AdS_4 \times N^{010}$ compactification
 of $D=11$ supergravity. This rescaled $N^{010}$ space is
 of particular interest for the $AdS_4/CFT_3$ correspondence, and its
 $SU(3) \times SU(2)$ isometric realization  is essential for the
 $OSp(4|3)$ classification of the Kaluza-Klein modes.

\end{abstract}
\vskip 4cm \noi \hrule \vskip .2cm \noi{\small e-mail:
castellani@to.infn.it}
\end{titlepage}
\newpage
\setcounter{page}{1}
\section{Introduction}

Coset manifolds are a natural generalization of group manifolds,
and play an important role in supergravity and superstring
compactifications, and in the recent AdS/CFT correspondences
\cite{adscft}. Indeed several of these correspondences have been
investigated in the context of  compactifications of supergravity
theories on anti-de Sitter spaces times ``internal" coset spaces
$G/H$. Many results of the 80's have been
reinterpreted and extended in the AdS/CFT framework, which has prompted
in particular a renewed interest in Kaluza-Klein mass
spectra of the $AdS \times G/H$ supergravity compactifications.
For an exhaustive list of references on this subject we refer
to the introduction of \cite{stiefel}. Here we will cite only the
papers dealing with $N^{pqr}$ spaces (see later).

In this note we generalize some formulae of the Riemannian
geometry of coset manifolds to include interesting cases, as the
$N^{010}$ spaces in the manifest $SU(3) \times SU(2)$ invariant
formulation. The general formulas of ref. \cite{CRWcosets,CDFbook} are
valid only for diagonal Killing metric, and need to be extended
for nondiagonal Killing metrics. While it is true that the Killing
metric can always be made diagonal by a redefinition of the group
generators, it may happen that the G/H structure we want to obtain
prevents such a redefinition, and that we must live with a
nondiagonal Killing metric. For the geometry of the $N^{pqr}$
coset spaces, and their use in $D=11$ supergravity compactifications,
we refer to the original papers \cite{CRNpqr,LCNpqr,PPN010}.
Recent developments using $N^{pqr}$ geometry to derive
Kaluza-Klein mass spectra and test $AdS_4/CFT_3$
correspondence are found in \cite{PTN010,FGTN010,ARN010}.

We give now a short review of coset space geometry, beginning with
a few definitions. A metric space is said to be
homogeneous if it admits as an isometry the transitive action of a
group G, transitive meaning that any two points of the space are
connected via the group action. For example the unit sphere $S^2$
in $\Rb^3$ is isometric under the transitive action of $SO(3)$.
The subgroup $H$ of $G$ which leaves a point $X$ fixed is called
the isotropy subgroup. Because of the transitive action of $G$,
any other point $X'=gX~(g \in G,g \notin H)$ is invariant under a
subgroup $gHg^{-1}$ of $G$ isomorphic to $H$. In the $S^2$ example
any point remains fixed under $SO(2)$ rotations around the axis
passing through that point, so that $SO(2)$ is the isotropy
subgroup.

It is natural to label the points $X$ of a homogeneous space by
the parameters describing the $G$ - group element which carries a
conventional $X_0$ (the origin) into $X$. However these parameters
are redundant: there are infinitely many group elements $g$ such
that $X = gX_0$, due to $H$ - isotropy. Indeed if $g$ carries
$X_0$ into $X$, any other $G$ element of the form $gH$ does the
same, since $HX_0=X_0$. We are then led to characterize the points
of a homogeneous space by the cosets $gH$.

A homogeneous space is therefore a coset space $G/H$, i.e. the set
of equivalence classes of elements of $G$, where the equivalence
is defined by right $H$ multiplication ($g \sim g'$ if $g=g'h$,
with $g,g' \in G$ and $h \in H$). Thus the two-sphere $S^2$ can be
considered as the coset space $SO(3)/SO(2)$. In general for an
$n$-sphere $S^n = SO(n+1)/SO(n)$. The action of an element $g' \in
G$ on the coset $gH$  is simply given by the coset $g'gH$.

Taking $G$ to be a Lie group (as in our $S^2$ example), we obtain
coset {\sl manifolds}, endowed with a Riemannian structure as we
will discuss. The Lie algebra of $G$ can be split as:
 \eq
 \mathbb{G}=\mathbb{H}\oplus\mathbb{K} \label{HKsplit}
 \en
where $\mathbb{H}$ is the Lie algebra of $H$ and $\mathbb{K}$
contains the remaining generators, called ``coset generators". The
structure constants of G are defined by:
 \eqa
 & &[H_i,H_j]=C_{ij}^{~~k} H_k ~~~~~~~~~~~~~~~~~~H_i \in \mathbb{H}\nonumber\\
 & &[H_i,K_a]=C_{ia}^{~~j} H_j + C_{ia}^{~~b}K_b~~~~~K_a \in \mathbb{K}
 \nonumber\\
 & &[K_a,K_b]=C_{ab}^{~~j}H_j+C_{ab}^{~~c}K_c
 \ena
where the index conventions are obvious.

As discussed in ref. \cite{pvntrieste} (p. 251), whenever $H$ is
compact or semisimple (even if $G$ is not compact) one can always find a set
of $K_a$ such that the structure constants $C_{ia}^{~~j}$ vanish.
In that case the $\mathbb{G}=\mathbb{H}+\mathbb{K}$ split, or equivalently the coset space $G/H$
is said to be {\sl reductive}. For this reason we will deal in this
note only with reductive coset spaces. Another important observation
is that when $G/H$ is reductive the structure constants $C_{ia}^{~~b}$
can always be made antisymmetric in $a,b$ by an appropriate
redefinition $K_a \rightarrow N_a^{~b} K_b$. The proof is simple:
any representation of a compact $H$ can be made unitary by a suitable
change of basis. Since the  $C_{ia}^{~~b}$ generate a real
representation of $H$ (namely the coset representation), this
representation can be made orthogonal, and consequently the
$C_{ia}^{~~b}$ antisymmetric \cite{pvntrieste,pvnleshouches}.

 An element $g$ of $G$ is
specified by $dimG$ continuous parameters, the Lie group
coordinates. For example we can exponentiate  the Lie algebra as:
 \eq
 g = \exp  [y^a K_a] \exp [x^i H_i] \label{expKexpH}
 \en
The $G$ coordinates are $y^a,x^i$. It is clear that the
cosets $gH$ are characterized by a subset of the group
coordinates, i.e. by the $dimG-dimH$ parameters $y^a$
corresponding to the $K_a$ generators.

Each coset, labeled by the $y$ parameters, can be mapped into an
element $L(y)$ of $G$,  the {\sl coset representative}. For
example one can choose as coset representative:
 \eq
  L(y) = \exp  [y^a K_a] \label{repchoice}
 \en

The whole geometry of $G/H$ can be constructed in terms of coset
representatives.

 Under left multiplication by a generic element
$g$ of $G$, $L(y)$ is in general carried into an element of $G$
belonging to another equivalence class, with representative
element $L(y')$, i.e. into an element of the form $L(y')h$:
 \eq
gL(y)=L(y')h,~~~h \in H \label{gL}
\en
where $y'$ and $h$ depend on $y$ and $g$, and on the way of
choosing representatives.  For example, using the representative
choice (\ref{repchoice}), $gL(y)$ loses the $\exp(yK)$ form but can be
expressed (as any element of $G$) as $\exp(y'K)\exp(xH)$ or
$L(y')\exp(xH)$. It is clear that the geometry of $G/H$ must be
insensitive to the particular representative choice, and indeed
 this is so (see later).

\section{Vielbeins, invariant metric, $H$-connection on $G/H$}

Consider the 1-form:
 \eq
  V(y)=\Linv (y) dL(y)
  \en
generalizing the left-invariant 1-form $g^{-1} dg$ of group
manifolds. $V(y)$ is Lie algebra valued and can expanded on the
$G$ generators:
 \eq
 V(y)=V^a(y)T_a + \Omega^i(y) T_i
\en
The 1-form $V^a(y)=V_{\al}^{a}dy^{\al}$ is a covariant frame
(vielbein)  on $G/H$ and $\Omega_{\al}^{~i}dy^{\al}$ is called the
$H$-connection.

Under left multiplication by a constant $g \in G$, the one-form
$\Linv dL$ is not invariant, but transforms as:
 \eq
 V(y')=h \Linv (y) g^{-1} d (gL(y)h^{-1})=hV(y)h^{-1}+h d h^{-1}
 \label{Vtransf}
 \en
 In particular its projection on the coset generators yields the
 transformation rule of the vielbein:
 \eq
 V^a(y')=(hV(y)h^{-1})^a=V^b(y) D_b^{~a}(h^{-1})
 \label{vielbeintransf}
 \en
where the adjoint representation $D_A^{~B}$ is defined by
$g^{-1}T_A g = D_b^{~a}(g)T_B$.

The infinitesimal form of (\ref{gL}) is obtained by taking:
 \eq
 g=1 + \epsi^A T_A
 \en
 Consequently, also the induced  $h$ transformation
  is infinitesimal:
 \eq
  h = 1 - \epsi^A W^i_{A}(y)T_i \label{compensator}
  \en
  and the shift in $y$ is proportional to $\epsi^A$:
  \eq
 y'^{\al}=y^a+\epsi^A K_A^{~\al}(y) \label{yshift}
 \en
The $y$ dependent matrix $ W^i_{A}(y)$ defined in
(\ref{compensator}) is called the $H$-compensator, and the
$y$-dependent differential operator
 \eq
 K_A(y) \equiv K_A^{~\al}(y) {\partial \over \partial y^{\al}}
 \en
 is the Killing vector on $G/H$ associated to the $G$-generator
 $T_A$. The explicit expressions for the $H$-compensator and the
 Killing vectors are simply obtained by rewriting the
 transformation rule (\ref{gL}) for infinitesimal $g$:
 \eq
 T_A L(y)=K_A (y) L(y) - L(y) T_i W_A^{~i}(y)
 \en
 After multiplying on the left by $\Linv  (y)$ and projecting
 on the $K$ and $H$ generators we find:
  \eqa
  & & K_A^{~\al}(y)=D_A^{~a} (L(y)) V_a^{\al} (y)\\
  & &  W_A^{~i}(y)=\Omega_{\al}^{~i}(y)K_A^{~\al}(y) -
  D_A^{~i}(L(y))
  \ena
where $V_a^{\al} (y)$ is defined as the inverse of the $G/H$
vielbein $V_{\al}^{a}$.

The infinitesimal form of the vielbein transformation
(\ref{vielbeintransf}) reads:
 \eqa
& & V^a(y+\de y) - V^a(y)=-\epsi^AW_A^{~i}(y)C_{ib}^{~~a}V^b(y)
\label{vielbeininf} \\
 & & \de y^{\al}=\epsi^A K_A^{~\al}(y)
 \ena
 easily derived by observing that the $C_{iB}^{~~A}$ are the
 generators of the adjoint representation of $H$, and
 $C_{ij}^{~~a}=0$.

 For reductive algebras $C_{ib}^{~~a}$ can be made antisymmetric in
 $a,b$: then eq. (\ref{vielbeininf}) implies that the left action
 of $G$ on $V^a(y)$ is equivalent to an $SO(N)$ rotation on
 $V^a(y)$  ($N=dim ~G/H$). Then the ``natural" coset metric
 \eq
g_{\al\be}=\de_{ab}V_{\al}^{a}V_{\be}^{b} \label{natmetric}
\en
is invariant under the left action of $G$. Another $G$
left-invariant metric is obtained by replacing the Kronecker delta
in (\ref{natmetric}) with the Killing metric $\ga_{AB} \equiv
C_{AD}^{~~~C} C_{BC}^{~~~D}$ restricted to $G/H$
 \eq
g_{\al\be}=\ga_{ab}V_{\al}^{a}V_{\be}^{b} \label{kmetric}
\en
Notice that both these invariant metrics are insensitive to the
choice of coset representative. Indeed replacing $L(y)$ by $L(y)h$
just rotates the vielbein as in (\ref{vielbeininf}).
\sk
Transformations that leave the metric invariant are called
 isometries. From the preceding discussion we know that
the isometries of $G/H$ manifolds include the left action of
$G$. However one can study also the right action of $G$ on the
coset representative:
\eq
L(y)g=L(y')h
\en
Then one finds that $N(H)/H$ is the right isometry group of $G/H$,
where $N(H)$ is the normalizer of $H$ in $G$, i.e. the set of
elements $g \in G$ such that $gHg^{-1}=H$. One is led to conclude
that the full isometry group  of $G/H$ must include
 $G \times N(H)/H$:
however this is not always true, as argued in ref.s
 \cite{CRWcosets,CDFbook}. Some left $U(1)$ Killing vectors may coincide with some
right $U(1)$ Killing vectors: then the actual isometry is reduced to
$G' \times N(H)/H$ where $G=G' \times $ (common $U(1)$ - factors).

\section{Rescaled Riemann connection and curvature}

In general the two metrics (\ref{natmetric}),(\ref{kmetric})
of the preceding Section are not
the only $G$-invariant metrics on $G/H$. As discussed in
various ref.s (see for example \cite{CRWcosets,pvntrieste,CDFbook})
whenever $C_{ia}^{~~~b}$ is block diagonal in some subspaces
$S_1,S_2,...$ of $\mathbb{K}$, then the vielbeins spanning these
subspaces can be independently rescaled without loss of left $G$
symmetry. This is easily understood from the transformation rule
(\ref{vielbeininf}), which remains unaltered when the vielbeins
belonging to the same subspace $S_i$ are rescaled by a common
parameter $r_i$. Therefore the number of rescaling parameters,
i.e. the number of parameters necessary to specify the particular
$G$-invariant metric, is equal to the number of irreducible blocks
of $C_{ia}^{~~~b}$. This matrix describes how $H$ acts on the
subspace $\mathbb{K}$: if it acts irreducibly, the coset is called isotropy
irreducible, and only the trivial rescaling $V^a \rightarrow r V^a$
(same $r$ for all $V^a$) is $G$-symmetric. If $G/H$ is isotropy
reducible, we have an independent parameter for each irreducible
subspace $S_i$. These rescalings must be real and nonsingular, but are
otherwise unconstrained. We derive now the expressions for the
Riemann connection and the curvature corresponding to the rescaled
vielbeins.
\sk
Recall the Cartan-Maurer equation for the one-form $V=\Linv dL$:
 \eq
 dV+V \we V=0
 \en
 which follows immediately from the definition of $V$. In components
 the Cartan-Maurer equation becomes:
 \eqa
 & & dV^a + {1 \over 2} C_{bc}^{~~a} V^b \we V^c + C_{bi}^{~~a} V^b
 \we \Omega^i =0 \\
 & & d\Omega^i + {1 \over 2} C_{ab}^{~~i} V^a \we V^b +
  C_{jk}^{~~i} \Omega^j \we \Omega^k=0
  \ena
  After a rescaling
  \eq
  V^a \rightarrow r_a V^a
  \en
the above equations become:
 \eqa
 & & dV^a + {1 \over 2} {r_b r_c\over r_a} C_{bc}^{~~a} V^b \we V^c
 + {r_b\over r_a} C_{bi}^{~~a} V^b
 \we \Omega^i =0 \label{dV}\\
 & & d\Omega^i + {1 \over 2} r_a r_b  C_{ab}^{~~i} V^a \we V^b +
  C_{jk}^{~~i} \Omega^j \we \Omega^k=0 \label{dOmega}
  \ena
For a $G$-symmetric rescaling we can replace ${r_b\over r_a}
C_{bi}^{~~a}$ by $ C_{bi}^{~~a}$ in the first equation.
\sk
The flat coset metric will be chosen in the following to be
 $\eta_{ab}=\eta^{ab}= - \de_{ab}$,
yielding a $G$-invariant metric $g_{\al\be} = \eta_{ab} V_{\al}^{~a}
V_{\be}^{~b}$.
\sk
A (torsionless) connection $B^a_{~b}$ on $G/H$ can be defined
 by the equation
 \eq
 dV^a + B^a_{~b} \we V^b =0 \label{zerotorsion}
 \en
 Combining (\ref{zerotorsion}) with (\ref{dV}) yields
  \eq
 B^a_{~b}= - {1 \over 2} {r_b r_c\over r_a} C_{bc}^{~~a} V^c
 - C_{bi}^{~~a} \Omega^i + K_{bc}^{~~a} V^c \label{conn}
 \en
where $K_{bc}^{~~a}$ is symmetric in $b,c$, and is determined by the
requirement that $B^a_{~b}$ be antisymmetric in $a,b$ (Riemann
connection):
\eq
B^a_{~c} \eta^{cb} = - B^b_{~c} \eta^{ca}
\en
Then:
 \eq
  K_{bc}^{~~a}= {r_a \over 2} \eta^{ad} \left({r_c\over
r_b} \eta_{be}C_{dc}^{~~e}+{r_b\over r_c} \eta_{ce}C_{db}^{~~e}
\right)
\en
and the antisymmetric connection is given by:
\eq
 B^a_{~b}= {1\over 2}\left(- {r_b r_c\over r_a} C_{bc}^{~~a} +
  {r_a r_c\over r_b}
 \eta_{bg} C_{dc}^{~~g} \eta^{ad} + {r_a r_b \over r_c} \eta_{cg}
 C_{db}^{~~g} \eta^{ad}\right) V^c - C_{bi}^{~~a} \Omega^i
 \label{connection}
\en
This connection is $G$-invariant, meaning that parallel transport
commutes with the $G$-action. Indeed the most general form of a
$G$-invariant connection on $G/H$ is given by
\eq
B^{a}_{~b}(y)= C_{ib}^{~~a}\Omega^{i}(y) +
J_{c~b}^{~a}V^c (y)
\en
where $J_{d~b}^{~a}$ is an invariant tensor of the subgroup $H$
\cite{pvnleshouches}, i.e. $\de J_{c~b}^{~a}= C_{i~c}^{~d}
J_{d~b}^{~a}- C_{i~d}^{~a} J_{c~b}^{~d}+ C_{i~b}^{~d}
J_{c~d}^{~a}=0$.
  The connection in (\ref{connection}) has this
form, and it is not difficult to prove that the term multiplying
$V^c$ is $H$-invariant. In fact each of the three terms
within parentheses in (\ref{connection}) is $H$-invariant,
as one can show by using Jacobi identities and ${r_a \over
r_b}~C_{ia}^{~~b}= C_{ia}^{~~b}$.
\sk
The Riemann curvature is defined in terms of $B^a_{~b}$ by:
\eq
R^{a}_{~b} \equiv dB^{a}_{~b} + B^a_{~c} \we B^{c}_{~b} \equiv
R^a_{~b~de} V^d \we V^e \label{curvaturedef}
\en
Substituting the connection (\ref{connection}) in the curvature
formula, using the Cartan-Maurer equations (\ref{dV}) and
(\ref{dOmega}) for $dV^a$ and $d\Omega^i$, and Jacobi identities for
products of structure constants, we determine the curvature
components:
\eq
 R^a_{~b~de} = {1\over 4} {r_d r_e \over r_c} \Cb_{bc}^{~~a} C_{de}^{~~c}
 + {1\over 2} r_d r_e~C_{bi}^{~~a} C_{de}^{~~i} + {1\over 8}
 \Cb_{cd}^{~~a} \Cb_{be}^{~~c} - {1 \over 8} \Cb_{ce}^{~~a}
 \Cb_{bd}^{~~c}
 \en
with
\eq
 \Cb_{bc}^{~~a} \equiv {r_b r_c\over r_a}~ C_{bc}^{~~a} - {r_a r_c\over r_b}~
 C_{ac}^{~~b} - {r_a r_b \over r_c}~ C_{ab}^{~~c}
 \en
These formulae generalize those of ref. \cite{CRWcosets,CDFbook} (holding
only for diagonal Killing metric) and those of \cite{pvntrieste}
(for unrescaled vielbeins). The connection $B$ allows the
definition of a covariant derivative $D$. For example the
zero-torsion condition can be written as $DV^a=0$. Taking the
exterior derivative of the zero-torsion condition
(\ref{zerotorsion}) and of the  curvature definition
(\ref{curvaturedef}) yields the Bianchi identities:
 \eqa & &
R^a_{~b}\we V^b=0 \label{Bianchitorsion}\\ & &
dR^a_{~b}+R^a_{~c}\we B^c_{~b} - B^a_{~c} \we R^c_{~b} \equiv
DR^a_{~b} = 0 \label{Bianchicurvature}
 \ena
 What are the symmetries
of the indices in the curvature components $R^a_{~b~cd}$ ?
Antisymmetry in $a,b$, and in $c,d$ is manifest. Furthermore, from
(\ref{Bianchitorsion}) we deduce:
 \eq R^a_{~b~cd} V^b \we V^c \we
V^d =0 ~~\Rightarrow ~~ R^a_{~[b~cd]}=0
\en
i.e. the cyclic identity. Using all these index symmetries one can
also show that $\R{ab}{cd} = \eta^{be} R^a_{~e~cd}$ is symmetric
under $ab \leftrightarrow cd$ interchange.

\section{The geometry of the $N^{010}$ coset manifolds}

We apply here the formulae of the preceding Section to the
coset manifolds $N^{010}$. These coset spaces are a special
case in the class of the $N^{pqr}$ coset spaces
defined by the quotient:
\eq
N^{pqr}={G\over H} = {SU(3) \times U(1) \over U(1) \times U(1)}
\en
where the $p,q,r$ are integer and coprime, and specify how
 the two $U(1)$ generators $M,N$ of $H$ are embedded into
$G$:
\eqa
& & M = -{ \sqrt{2} \over RQ} \left( {i\over 2}rp \sqrt{3} \lambda_8 +
{i\over 2}rq\lambda_3 -{i\over 2}(3p^2 +q^2)Y \right) \\
& &N= -{1\over Q} \left( -{i\over 2}q\lambda_8 + {i\over 2}p\sqrt{3}
\la_3 \right)\\
& & Z= -{1\over R} \left( {i\over 2}p\sqrt{3}\la_8 + {i\over 2}
q\la_3 +i r Y \right)
\ena
with
\eq
R=\sqrt{3p^2+q^2 +2r^2},~~~~Q=\sqrt{3p^2+q^2}
\en
and $Z$ is the remaining $U(1)$ generator in the coset. The
generators of $G=SU(3) \times U(1)$ are taken to be $-{i\over 2}
\la$ and $-{i\over 2}Y$, $\la$ being the Gell-Mann matrices. For a
detailed account of the geometry of these $N^{pqr}$ coset
manifolds we refer to the original papers \cite{CRNpqr,LCNpqr},
where symmetric rescalings, connection and curvature are given
explicitly. The cosets $N^{pqr}$ for $p=0,q=1,r=0$
have as isometry group $SU(3) \times SU(2)$ (coming from $G \times
N(H)/H$). As already observed in \cite{CRNpqr}, the $N^{010}$
cosets can also be realized as:
 \eq
  N^{010}={SU(3) \times SU(2) \over
SU(2) \times U(1)} \label{N010}
\en
where the $SU(2)$ in the denominator is diagonally embedded in
$G = SU(3) \times SU(2)$. In this formulation the full isometry
of $N^{010}$ comes from the left action of $G$. We now study the
geometry of $N^{010}$ realized as in (\ref{N010}).
\sk

The generators of $SU(3)$ and $SU(2)$ are taken respectively to be
$-{i\over 2} \la$ and $-{i\over 2} \tau$, $\la$ being the
Gell-Mann matrices and $\tau$ the Pauli matrices, with commutation
relations:
 \eq
 [\la_i,\la_j]=2i~f_{ijk}\la_k,~~~[\tau_m,\tau_n]=2i~\epsilon_{mnr}\tau_r
 \en
 where the nonvanishing components of the completely antisymmetric
 structure constants $f_{ijk}$ are
 $f_{123}=1,f_{147}=f_{165}=f_{246}=f_{257}=f_{345}=f_{376}={1\over
 2},f_{458}=f_{678}={\sqrt{3}\over 2}$.

 The $U(1)$ in the denominator of (\ref{N010}) is
given by the hypercharge $-{i\over 2} \la_8$. Thus the
$\mathbb{H}+\mathbb{K}$ generators are:
\sk
\noi $\mathbb{H}$-generators:
 \eq
H_N= -{i\over 2}\la_8,~~H_i=-{i\over 2}
 (\la_1+\tau_1,\la_2+\tau_2,\la_3+\tau_3)
 \en

\noi $\mathbb{K}$-generators:
 \eq
 K_a=-{i\over 2}
 (\la_1-\tau_1,\la_2-\tau_2,\la_3-\tau_3),~~K_{\dot{A}}=
 -{i\over 2}(\la_4,\la_5),~~K_{\udot{A}}=
 -{i\over 2}(\la_6,\la_7)
 \en

The $\mathbb{H}+\mathbb{K}$ basis is reductive, and the
nonvanishing structure constants are:

 \eq
C_{HK}^{~~K}:~~~C_{N\dot{A}}^{~~\dot{B}}={\sqrt{3}\over 2}
 \epsilon_{\dot{A}\dot{B}},~~C_{N\udot{A}}^{~~\udot{B}}={\sqrt{3}\over 2}
 \epsilon_{\udot{A}\udot{B}},~~C_{ia}^{~~b}=\epsilon_{iab},
 ~~C_{iA}^{~~B}=f_{iAB}
 \en
 \eqa
& &  C_{KK}^{~~K}:~~~C_{aA}^{~~B}=f_{aAB},~~C_{AB}^{~~c}=
  {1\over 2}f_{ABc} \\
& &  C_{KK}^{~~H}:~~~C_{ab}^{~~i}=\epsilon_{abi},
  ~~C_{AB}^{~~N}=f_{AB8},~~C_{AB}^{~~i}=
  {1\over 2}f_{ABi}\\
& & C_{HH}^{~~H}:~~~C_{ij}^{~~k}=\epsilon_{ijk}
 \ena
 with $A=(\dot{A},\bar{A})=4,5,6,7$.
The Killing metric in this basis is diagonal on the coset
directions and on the {\small H} directions:
 \eq
 \ga_{ab}=-5 \de_{ab},~~\ga_{AB}=-3\de_{AB}
 \en
 \eq
 \ga_{NN}=3,~~\ga_{ij}=-5 \de_{ij}
 \en
 and has nondiagonal components along {\small HK} directions:
 \eq
 \ga_{ia}=\ga_{ai}=-\de_{ai}
 \en
 By inspection of the $C_{HK}^{~~K}$ structure constants we see
 that these are antisymmetric in the two coset indices, and that
 there are two isotropy-irreducible subspaces, spanned
 respectively by the vielbeins $V^a$ ($a=1,2,3$) and $V^A$
 ($A=4,5,6,7$). We can therefore construct $G$-invariant metrics
 depending on two independent rescaling parameters, $\al=r_a$ and $\be=r_A$.
Applying the general formulae of the preceding Section we find the
Riemann connection 1-form :
 \eqa
 & &B^a_{~b}=-\epsilon_{bia} \Om^i\\
 & &B^a_{~\dot{B}}=-{\be^4\over
 8\al}\left(\de_{a1}\epsilon_{\dot{B}\udot{A}}V^{\udot{A}}+
 \de_{a2}\de_{\dot{B}\udot{A}}V^{\udot{A}}+
 \de_{a3}\epsilon_{\dot{B}\dot{A}}V^{\dot{A}}\right)\\
 & &B^a_{~\udot{B}}={\be^4\over
 8\al}\left(-\de_{a1}\epsilon_{\udot{B}\dot{A}}V^{\dot{A}}+
 \de_{a2}\de_{\udot{B}\dot{A}}V^{\dot{A}}+
 \de_{a3}\epsilon_{\udot{B}\udot{A}}V^{\udot{A}}\right)\\
 & &B^{\dot{A}}_{~\dot{B}}={1\over 8\al}\left[(-4\al^2+\be^2)V^3-
 4\al(\sqrt{3} \Om^8 + \Om^{11})\right]
 \epsilon_{\dot{A}\dot{B}}\\
 & &B^{\udot{A}}_{~\udot{B}}={1\over 8\al}\left[(4\al^2-\be^2)V^3-
 4\al(\sqrt{3} \Om^8 - \Om^{11})\right]
 \epsilon_{\udot{A}\udot{B}}\\
 & &B^{\dot{A}}_{~\udot{B}}={1\over 8\al}\left[(-4\al^2+\be^2)
 (\de_{\dot{A}\udot{B}}V^2+\epsilon_{\dot{A}\udot{B}}V^1)-
 4\al(\de_{\dot{A}\udot{B}}\Om^{10}+ \epsilon_{\dot{A}\udot{B}}
 \Om^{9})\right]
 \ena
and the corresponding Riemann curvature components:
 \eqa
 & & \R{ab}{cd}=\al^2 \de^{ab}_{cd},~~\R{ab}{\dot{A}\dot{B}}=
  {1\over 32} \ga \de^1_{[a} \de^2_{b]}
  \epsilon_{\dot{A}\dot{B}},~~\R{ab}{\udot{A}\udot{B}}=
  {1\over 32} \ga \de^1_{[a}
  \de^2_{b]}\epsilon_{\udot{A}\udot{B}} \label{riemannfirst}\\
  & & \R{ab}{\dot{A}\udot{B}}= {1\over 32} \ga  (-\de^1_{[a}
  \de^3_{b]}\de_{\dot{A}\udot{B}}+\de^2_{[a}
  \de^3_{b]}\epsilon_{\dot{A}\udot{B}}) \\
  & & \R{a\dot{A}}{b\dot{B}}={\be^4\over 128\al^2} \de^a_b
  \de^{\dot{A}}_{\dot{B}}+
 {1\over 64} \ga~ \de^1_{[a} \de^2_{b]}
  \epsilon_{\dot{A}\dot{B}}\\
  & &  \R{a\udot{A}}{b\udot{B}}={\be^4\over 128\al^2} \de^a_b
  \de^{\udot{A}}_{\udot{B}}-
 {1\over 64} \ga ~\de^1_{[a} \de^2_{b]}
  \epsilon_{\udot{A}\udot{B}}\\
  & & \R{a\dot{A}}{b\udot{B}}= - {1\over 64} \ga ~\de^1_{[a} \de^3_{b]}
  \de^{\dot{A}}_{\udot{B}},~~\R{a\udot{A}}{b\dot{B}}= {1\over 64}
   \ga ~ \de^1_{[a} \de^3_{b]} \de^{\udot{A}}_{\dot{B}}\\
   & & \R{\dot{A}\dot{B}}{\dot{C}\dot{D}}=
   \be^2 \left( 1-{3\over 64} {\be^2\over \al^2} \right)
   \de^{\dot{A}\dot{B}}_{\dot{C}\dot{D}},
   ~~\R{\udot{A}\udot{B}}{\udot{C}\udot{D}}=
    \be^2 \left( 1-{3\over 64} { \be^2\over  \al^2} \right )
    \de^{\udot{A}\udot{B}}_{\udot{C}\udot{D}}\\
   & &\R{\dot{A}\dot{B}}{\udot{C}\udot{D}}={ \be^2 \over 2}
   \de^{\dot{A}\dot{B}}_{\udot{C}\udot{D}},~~
   \R{\dot{A}\udot{B}}{\dot{C}\udot{D}}= {\be^2\over 8} \left[
  \left( 1-{3\over 16}{ \be^2\over  \al^2} \right )
 \de^{\dot{A}}_{\dot{C}} \de^{\udot{B}}_{\udot{D}}+
 \epsilon_{\dot{A}\dot{C}} \epsilon_{\udot{B}\udot{D}} \right]
 \label{riemannlast}
 \ena
with $\ga \equiv  \be^2 (8-\be^2/\al^2)$. The Ricci tensor is:
 \eqa
& & R_{ab}=\left(\al^2+{1\over 32}{\be^4\over  \al^2}\right)
\de_{ab} \label{ricci1}\\
 & & R_{AB}={3 \over 4} \be^2 \left( 1 -{1 \over 16} {\be^2 \over
 \al^2} \right) \de_{AB} \label{ricci2}
 \ena

 {\bf Note 4.1} : only the squares of the rescalings appear
 in the curvatures. On the other hand the connection depends on $\al$
 and $\be^2$: the sign of $\be$ has therefore no influence on the geometry,
  whereas different signs of $\al$ yield different
 spaces.

\section{$AdS_4 \times N^{010}$ as compactification of $D=11$ supergravity}

 As observed in the early eighties \cite{FR}, a nontrivial solution of the
 $D=11$ supergravity field equations is given by setting the
 gravitino curvature to zero, and taking the bosonic curvatures
 as:
 \eq
 R_{mn} = - 24 e^2 \de_{mn},~~R_{ab}= 12 e^2
 \de_{ab},~~F_{mnpq}=e\epsilon_{mnpq}
 \en
all other curvature components vanishing. The indices $m,n,p,q$
run on  4-spacetime and $a,b$ on the internal $7$-dimensional
space; $R_{mn}$ and $R_{ab}$ are the corresponding Ricci
curvatures, in our conventions $R_{mn} = R^q_{~m~qn}$, and
$F_{mnpq}$ is the curl of the antisymmetric three-index tensor.

Then all spaces of the type $AdS_4 \times$ (7-dimensional Einstein
space) are a solution of the supergravity equations, Einstein
space meaning a Riemannian manifold with Ricci tensor proportional
to the metric.  A classification of all 7-dimensional $G/H$
Einstein manifolds was derived in the eighties in
\cite{CRWclassification}, thus providing a class of $D=11$
supergravity solutions (for their use in the more recent
 $G/H$ $M$-branes see \cite{GHMbranes}). The coset manifolds $N^{010}$ studied in
the preceding Section are part of this classification, although
they were studied as particular instances of the $N^{pqr}$ spaces, in
the $SU(3) \times U(1)$ - isometric formulation. Two inequivalent
Einstein metrics were found, and the corresponding Einstein spaces
were denoted by $N_I^{pqr}$ and $N_{II}^{pqr}$
\cite{CRNpqr,PPN010,LCNpqr}.
 \sk
What can we say about Einstein metrics in the $N^{010}$ cosets
discussed in this paper ? As easily seen from the expression of
the Ricci tensor in (\ref{ricci1}), (\ref{ricci2}) the rescalings
 \eq
  \al^2=4 e^2, ~~~\be^2=32 e^2  \label{rescalings1}
  \en
  or
  \eq
 \al^2={100 \over 9} e^2, ~~~\be^2={160 \over 9} e^2
 \label{rescalings2}
  \en
both bring the Ricci tensor in the Einstein form $R_{ab}= 12 e^2
 \de_{ab}$. We denote by $N_I^{010}$ and $N_{II}^{010}$
 the corresponding Einstein coset spaces, since these coincide
 with the $N_I^{pqr}$ and $N_{II}^{pqr}$ for
 $p=0,q=1,r=0$, as one can easily prove by comparing the Riemann
 curvatures.
 \sk
 Finally, we  can investigate the supersymmetry content of the
 $AdS_4 \times N$ compactifications. We recall that the independent
 supersymmetry charges preserving the $AdS_4 \times N$ vacuum are
 in 1-1 correspondence with the number of spinors $\eta$ satisfying
 the equation:
 \eq
 (d+{1\over 4} B^{ab} \Ga_{ab} - e V^a \Ga_a )\eta=0
 \label{susypsi}
 \en
 which is just the requirement that the supersymmetry variation of
 the gravitino vanishes in the $AdS_4 \times N$ background (see for ex.
 \cite{DNPphysrep,CDFbook}).
 The integrability condition for  (\ref{susypsi}) is
 \eq
  (\R{cd}{ab} + 4 e^2 \de^{cd}_{ab})\Ga_{cd} \eta =0
  \label{integpsi}
  \en
  Substituting into  (\ref{integpsi}) the Riemann tensor of eqs.
  (\ref{riemannfirst})-(\ref{riemannlast}) with the rescalings
 $\al^2=4 e^2, ~~~\be^2=32 e^2 $ yields four independent spinors
 $\eta$ satisfying (\ref{integpsi}), while for the rescalings $\al^2={100 \over 9} e^2, ~~~\be^2=
 {160 \over 9} e^2 $  only one spinor $\eta$ exists.
 Then one has to check whether these spinors also satisfy
 (\ref{susypsi}). Whereas the sign of $\al$ is irrelevant
 in the integrability condition (since the Riemann curvature does not
 depend on it), it becomes important in the supersymmetry variation
 (\ref{susypsi}), and we find the following:
 \eqa
 & & N^{010}_I:~~~\al=2e,~~~\be=\pm 4
 \sqrt{2}~e,~~~~~~~~N=3~\mbox{supersymmetry}\\
  & & {\tilde {N}}^{010}_{I}:~~~\al=- 2e,~\be=\pm 4
 \sqrt{2}~e,~~~~~~~~N=0~\mbox{supersymmetry}\\
  & & N^{010}_{II}:~~~\al=-{10\over 3}e,~\be=\pm {4 \over 3}
  \sqrt{10} ~e ,~~~N=1~\mbox{supersymmetry}\\
   & & {\tilde{N}}^{010}_{II}:~~~\al= {10\over 3}e,~\be=\pm {4 \over 3}
  \sqrt{10}~ e ,~~~~~~N=0~\mbox{supersymmetry}
  \ena
  where we have denoted by ${\tilde N}$ the spaces
  obtained  by reversing the orientation of $N^{010}$, i.e.
 by taking $V^a \rightarrow -V^a$ or equivalently $\al,\be
 \rightarrow -\al,-\be$. Thus
  changing signs in $\al$  is equivalent
  to reverse the orientation, since the sign of $\be$ has no
  influence on the geometry.

{\bf Note 5.1:} in Ref. \cite{PTN010} the $N^{010}_I $ space
corresponds to the rescaling $\al=-2e$, due to a sign difference
in the structure constants of $G$.

{\bf Note 5.2:} the Killing spinors satisfying eq. (\ref{susypsi})
are {\sl not} constant in the $SU(3) \times SU(2)/ SU(2) \times U(1)$
realization of the $N^{010}$ spaces. On the other hand
they are constant
in the $N^{pqr}=SU(3) \times U(1)/U(1) \times U(1)$ spaces of
\cite{CRNpqr}, where for $p=0,q=1$ the isometry is promoted to
$SU(3) \times SU(2)$ because of the right action of $N(H)/H=SU(2)$.


\vfill\eject
\end{document}